\documentclass[]{rstransa}%%%%where rstrans is the template name

\usepackage{scalerel}
\usepackage{tikz}
\usetikzlibrary{svg.path}
\usepackage{graphicx}	% Including figure files
\usepackage{amsmath}	% Advanced maths commands
\usepackage{amssymb}	% Extra maths symbols
\usepackage[english]{babel}
\usepackage{float}      % figure position
\usepackage{nccmath}
\usepackage{float}
\usepackage{appendix}
\usepackage{subcaption}
\usepackage{comment}

\usepackage{xcolor} % Package for colors
\titlehead{Research}

\begin{document}

%%%% Article title to be placed here
\title{Dust production through collisions between small bodies: an application to the G-ring arc}
\author{%%%% Author details
V. Lattari$^{1}$, R. Sfair$^{1,2,3}$,  P. B. Siqueira$^{1}$, and C. M. Schäfer$^{2}$}

%%%%%%%%% Insert author address here

\address{
$^{1}$ São Paulo State University (UNESP), School of Engineering and Sciences, 12516-410, Guaratinguetá, Brazil \\
$^{2}$ Institute of Astronomy and Astrophysics, University of T{\"u}bingen, Auf der Morgenstelle 10, 72076, Tübingen, Germany \\
$^{3}$ LESIA, Observatoire de Paris, Université PSL, CNRS, Sorbonne Université, 5 place Jules Janssen, 92190 Meudon, France
}

%%%% Subject entries to be placed here %%%%
\subject{astronomy, planetary rings}

%%%% Keyword entries to be placed here %%%%

%%%% Insert corresponding author and its email address}
\corres{V. Lattari\\
\email{victor.lattari@unesp.br}}

%%%% Abstract text to be placed here %%%%%%%%%%%%
\begin{abstract}
The G-ring arc of Saturn, confined by the 7:6 corotation eccentric resonance with Mimas,
is primarily composed of micrometric particles. These particles, significantly influenced by the
solar radiation pressure, are subject to rapid depletion. This study investigates a mechanism
for dust replenishment in the arc, specifically analyzing collisions between macroscopic bodies
and the satellite Aegaeon. Utilizing N-body and Smoothed particle hydrodynamics (SPH) simulations, we assess the dust generation
from these impacts, with a focus on the most likely collision parameters derived from the N-body
simulations. Our findings indicate that, while collisions among macroscopic bodies are inefficient
for dust production, impacts involving Aegaeon are substantially more effective, 
providing conservative lower limits for dust generation. 
This mechanism, in conjunction with the natural decay processes and continuous dust generation
from impacts, potentially keeps the arc population over thousands of
years with a possible variation in brightness.
\end{abstract}
%%%%%%%%%%%%%%%%%%%%%%%%%%%

%%%%%%%%%% Insert the texts which can accomdate on firstpage in the tag "fmtext" %%%%%

%\begin{fmtext}
%\end{fmtext}

%%%%%%%%%%%%%%% End of first page %%%%%%%%%%%%%%%%%%%%%
\maketitle
\section{Introduction}
\label{section-introduction}
In early 2004, the Cassini spacecraft identified two tenuous arcs in eccentric
resonances with Saturn's satellite, Mimas. The existence of these arcs is linked
to the satellites Anthe and Methone, and both are captured within the 10:11
and 14:15 corotation eccentric resonances (CER) with Mimas \cite{hedman_2007}.

Cassini images also revealed a denser region at the edge of the G ring, analogous
to the Anthe and Methone arcs. The G-ring arc has a radius of approximately
167,500 km and spans a length of 60$^\circ$ \cite{hedman_2007}. At the center of
this arc's resonance lies the small satellite Aegaeon, which measures about 240
meters in radius and has a mass of $3\times10^{10}$ kg, assuming a density of 0.5
$kg/cm^3$. Similar to the other two arcs, the G-ring arc is azimuthally confined due
to the 7:6 corotation eccentric resonance (CER) with Mimas
\cite{hedman_2010,el2011dynamics,el2014coupling}. For this resonance, the
librating angle $\varphi$ is expressed as

\begin{align}
\varphi = 7\lambda_{Mimas} -6\lambda_{Aegaeon}-\varpi_{Aegaeon}.
\end{align}

\noindent where $\lambda_{Mimas}$ and $\lambda_{Aegaeon}$ denote the mean longitudes of Mimas and Aegaeon, 
respectively, and $\varpi_{Aegaeon}$ is the longitude of the pericentre of Aegaeon.

The G-ring arc's total dust mass is approximately $10^6$ kg. The strong back-scattering
observed in the G-ring arc suggests micrometric dust particles constitute the primary
population. Aegaeon imparts significant perturbations on the orbital paths of
these dust particles, leading to frequent collisions with nearby material and the small
satellite itself \cite{rodriguez2021dynamical}. Given their size, these dust particles are
susceptible to various disturbing forces. Solar radiation pressure, along with
electromagnetic and plasma drag, can perturb their orbits and reduce their lifespan
\cite{hedman_2007, madeira2020}.

Solar pressure increases the eccentricity of the dust particles, reducing their
lifetime. These dust particles can survive a maximum of 30 years for 10 $\mu$m grains
and three years for 1 $\mu$m grain. This duration is significantly shorter, over
a hundred times, than the time taken for Aegaeon to repopulate the arc through
collisions with interplanetary particles. This scenario contrasts with that of the
Anthe and Methone arcs, where both satellites generate sufficient dust from
collisions between interplanetary particles and the satellites, effectively replenishing
the arc before solar pressure can deplete the dust. It is thus concluded that the
satellites replenish the dust population of their respective arcs more rapidly than
solar pressure can eliminate them \cite{sun2017dust, hedman_2010, madeira_2018}.

The equivalent normal areas of the arcs associated with Aegaeon, Anthe, and Methone
are approximately 50, 1.0, and 0.3 $km^2$, respectively, while the equivalent areas of 
Aegaeon, Anthe, and Methone are 0.07,
0.84, and 2.21 $km^2$, respectively. These values suggests that Anthe and its arc exhibit
comparable brightness levels, whereas Methone is brighter than its corresponding arc.
In these instances, the mass of other bodies within these rings is negligible. However,
the G-ring arc's equivalent normal area is nearly a thousand times larger than that of
Aegaeon, indicating the presence of additional macroscopic bodies in the arc to account
for the observed brightness \cite{hedman_2007, hedman_2010}.

The dust generation process in Aegaeon's arc differs markedly from that in Anthe's and
Methone's arcs. Aegaeon takes approximately 30,000 years to produce a comparable amount
of dust in its arc, a period three orders of magnitude longer than the time radiation
pressure needs to remove dust particles from the arc \cite{madeira_2018}. This significant
discrepancy highlights the necessity for an alternative mechanism to maintain a steady-
state dust particle population in Aegaeon's arc. This mechanism likely involves the combined
effects of the ring's mean radius, its brightness, and the presence of macroscopic bodies
within the G-ring arc \cite{hedman_2010}.

To estimate the total mass of these macroscopic bodies, the Cassini spacecraft's Low
Energy Magnetospheric Imaging Instrument (LEMMS) was utilized. LEMMS measured the electron
flux within the ring, and specifically in the G-ring's arc, it detected an energy reduction
ranging from 1 to 10 MeV, suggesting that other bodies are absorbing these electrons. This
energy drop corresponds to a total mass estimate of approximately $10^8$ to $10^{10}$ kg,
assuming a hypothetical ice composition with a density of 0.92 $g/cm^3$ \cite{hedman_2007,
hedman_2010}. It is also relevant to note that due to backward scattering, Cassini's cameras
would only detect objects with a minimum radius of one hundred meters, implying that the mass
observed by LEMMS is likely distributed among smaller bodies \cite{hedman_2010}.

The presence of macroscopic bodies ranging from 1 to 100 meters in radius within the G-ring arc influences
its dynamics and increases the chance of a collision with Aegaeon. In densely populated and
confined arcs, collisions can increase the libration amplitude and displace bodies from the
resonance. However, given Aegaeon's larger size compared to these other bodies, collisions may
reduce its eccentricity and libration without displacing it from the resonance \cite{hedman_2010}.
When massive bodies are trapped in corotation resonance, they exchange angular momentum and
energy during close encounters. Yet, the movement of one body towards corotation does not
necessarily result in the other body moving away from exact corotation, allowing post-encounter
trajectories in various directions \cite{a2019dynamics}.

According to \cite{hedman_2010}, the presence of Aegaeon and other bodies in the G-ring arc could
be attributed to super-catastrophic collisions, with Aegaeon being the larger of the colliding
bodies. An alternative hypothesis suggests that Aegaeon may have been captured during Mimas'
migration from the A ring \cite{araujo_2016, el2017derivation}. Further investigation is required
to determine the precise origins of Aegaeon and other bodies within the G-ring arc. These insights
contribute to our understanding of Saturn's ring system's dynamics and the various forces shaping
its structure.

This study focuses on examining collisions among macroscopic 
bodies in the G-ring's arc and their
interactions with Aegaeon, to estimate dust production through these events. The objective is to
determine whether these bodies can replenish the arc's dust and explore scenarios where dust
production balances the rate at which solar pressure removes dust particles from the arc. 

The article is structured as follows: Section \ref{section-numericalsetup} details the numerical 
setup for n-body simulations, while Section \ref{section-outcomecollisions} analyzes the collision 
distributions. Section \ref{section:sph} evaluates dust production through smoothed particle hydrodynamic 
simulations, and Section \ref{section-dustproduction} discusses dust evolution in the arc. 
The examination of Aegaeon's mass evolution is covered in Section \ref{section-massevolution}. 
Finally, our results are summarized and discussed in Section \ref{section-finalremarks}.

%%%%%%%%%%%%%%%%%%%%%%%%%%%%%%%%%%%%%%%%%%%%%%%%%%

\section{Numerical setup}
\label{section-numericalsetup}

In this section, we present the setup of the numerical simulations employed to
calculate collision parameters within the arc. This is necessary for determining
typical collision parameters, including velocity and impact parameter,
prevalent in the arc. These parameters will subsequently serve as initial
conditions for the Smoothed particle hydrodynamics (SPH) simulations, as
discussed in Section \ref{section:sph}.

We used the freely accessible modular N-body integrator \textsc{Rebound}
\cite{rebound}. We chose the non-symplectic high-precision integrator with an
adaptive time-step, \textsc{IAS15} \cite{IAS15}, and implemented additional
terms corresponding to the acceleration due to Saturn's non-sphericity.

To determine the initial conditions for our simulations, we used the physical
parameters and gravity coefficients for Saturn shown in Table \ref{tab:saturn},
while Table \ref{tab:mimas_aegaeon} presents the elements for the satellites. In
this work, we performed two different sets of simulations: the first including
the satellite Aegaeon and the second without it, to understand the effect of the
satellite on particles embedded in the arc. In all simulations, we included the
$J_2$, $J_4$, and $J_6$ coefficients to determine the arc positions
corresponding to the 7:6 resonances with Mimas.

\begin{table}[!ht]
	\centering
	\caption{Saturn's physical parameters \cite{hedman_2010}.}
	\label{tab:saturn}
	\begin{tabular}{lr} % four columns, alignment for each
		\hline
		Mass & 5.683$\times$10$^26$ \\ 
		Radius ($km$) & 60330 \\ 
        $J_2$  & 0.016290543820 \\
        $J_4$  & -0.000936700366  \\
        $J_6$  & 0.000086623065  \\
        \hline
	\end{tabular}
\end{table}

\begin{table}[!ht]
	\centering
	\caption{Osculating orbital elements of the satellites Mimas and Aegaeon corresponding 
 to the epoch 08/22/2008 UT 00:00:00, as described in \cite{less2019}}.
	\label{tab:mimas_aegaeon}
	\begin{tabular}{lll} % four columns, alignment for each
		\hline
		Orbital Elements & Aegaeon & Mimas \\ 
		\hline
        a  ($\times 10^5$ km) & 1.6803398728 & 1.8600466879 \\ 
        e ($\times 10^{-2}$)  & 0.3133178063 & 1.7245224226 \\
        I ($^\circ$)          & 0.0017328550 & 1.5641747812 \\ $\varpi$ ($^\circ$)   & 142.49114788 & 163.18023984 \\
        $\Omega$ ($^\circ$)   & 236.30175623 & 259.15258436 \\ $M$ ($^\circ$)        & 5.3481201068 & 197.73278953 \\	\hline
	\end{tabular}
\end{table}

To distribute into the 7:6 CER the particles that represent the 
macroscopic bodies,  
we utilized the algorithm described by \cite{renner_sicardy_2006}. 
This procedure compensates for the forced eccentricity 
$e_0 \sim (3/2)J_2 (R/r)^2$ induced by Saturn's oblateness \cite{borderies1994test}. 
Our approach involved an initial distribution of 10,000 test 
particles on circular orbits around the resonance balance point. 
By calculating the resonance angle for each particle, we identified that 
approximately 65\% become trapped in resonance. 
From this resonant population, we then randomly selected particles to incorporate into our subsequent simulations.

In the simulations, the bodies are assumed to be 20 m in radius and have
densities of either 0.5 or 0.9 g/cm$^3$. This choice reflects densities similar
to icy bodies observed in other Saturnian rings \cite{hedman_2007}, with those
having 0.5 g/cm$^3$ being considered as porous. 
The choice of 20 m radius bodies represents a balance between 
computational efficiency for N-body simulations and physical relevance, 
being below Cassini's detection threshold ($\sim 50$ m) \cite{hedman_2007}. 
While a truncated power law distribution would be physically motivated given 
the collisional evolution of the system, collisions between smaller 
bodies would likely be as inefficient at dust production as those we 
observed between 20 m objects, 
since relative velocities are expected to be of the same order of magnitude. These 
velocities and collision outcome will be presented in the next sections.

With these
parameters, the arc population contains approximately 1,200 macroscopic bodies
with a density of 0.9 g/cm$^3$, or about 2,000 bodies at 0.5 g/cm$^3$, to
account for the total estimated mass of $7\times10^9$ kg. This mass is
equivalent to the upper limit derived from the LEMMS instrument, as discussed in
Section~\ref{section-introduction}.

%%%%%%%%%%%%%%%%%%%%%%%%%%%%%%%%%%%%%%%%%%%%%%%%%%%%
%Outcome of the collisions
\section{Outcome of the collisions}
\label{section-outcomecollisions}
This section examines the velocity distribution and impact angles in collisions
from our simulations. The simulations were divided into two groups with
identical initial conditions for the bodies, differing only in the inclusion of 
Aegaeon. The analysis covers collisions among the bodies and
between the bodies and Aegaeon, quantifying typical velocities and impact
angles to estimate dust production in both scenarios.

\begin{figure}[!ht]
    \centering
    \includegraphics[width=0.49\columnwidth]{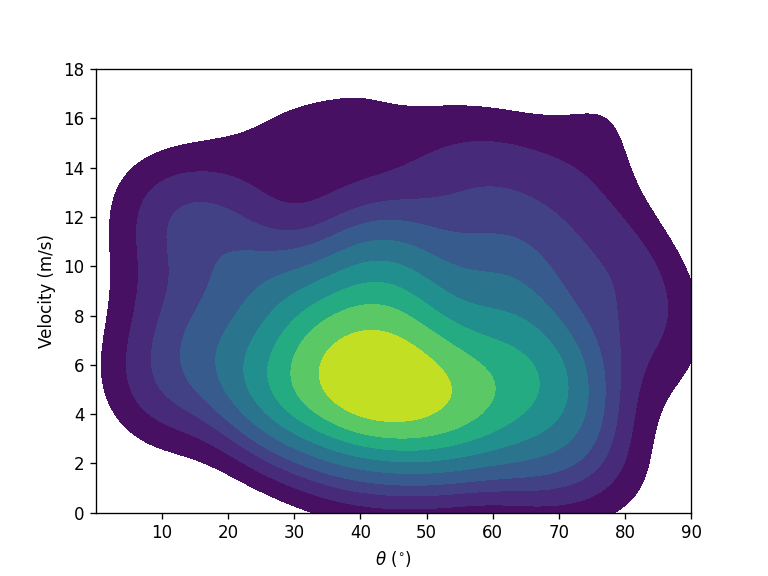}
    \includegraphics[width=0.49\columnwidth]{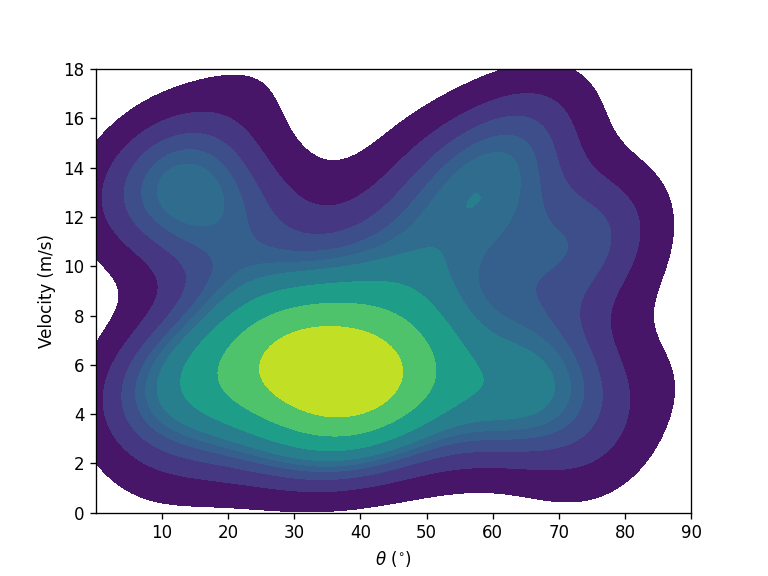}
    \caption{Relative velocities as a function of the impact angle ($\theta$) in collisions
    between a 20 m radius body and Aegaeon. The left panel depicts the simulation with a
    density of 0.5 g/cm$^3$, and the right panel with 0.9 g/cm$^3$. 
    The color
    map ranges from less likely (blue) to more likely (yellow).}
    \label{fig:collision_aegaeon}
\end{figure}

Figure \ref{fig:collision_aegaeon} presents a heatmap of relative velocity
versus impact angle for collisions between a 20 m radius body and Aegaeon. The
left panel shows simulations with 0.5 g/cm$^3$ body density, the right with 0.9
g/cm$^3$. Typical collision velocities are approximately 5 m/s, ranging from 1
to 15 m/s, with impact angles predominantly between 30 and 50 degrees. No
significant variance is observed between simulations with different body
densities, as density variations only affect the number of bodies in the
simulations.

\begin{figure}[!ht]
    \centering
    \includegraphics[width=0.49\columnwidth]{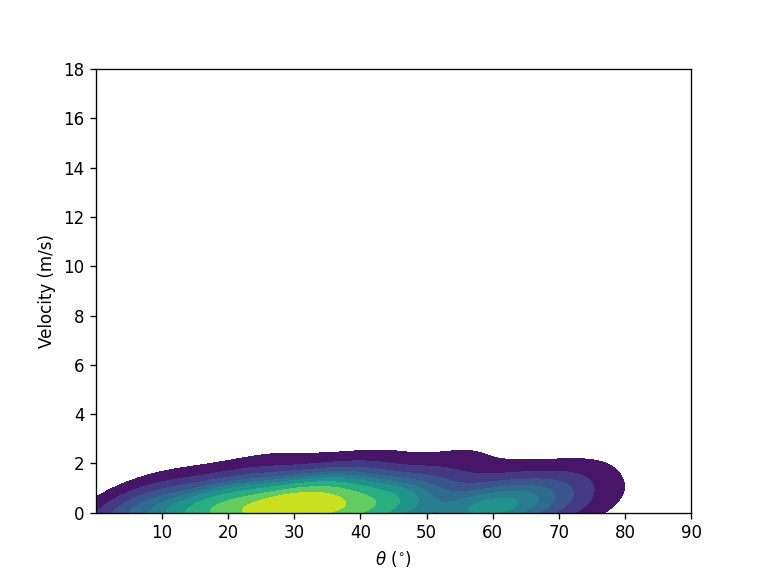}
    \includegraphics[width=0.49\columnwidth]{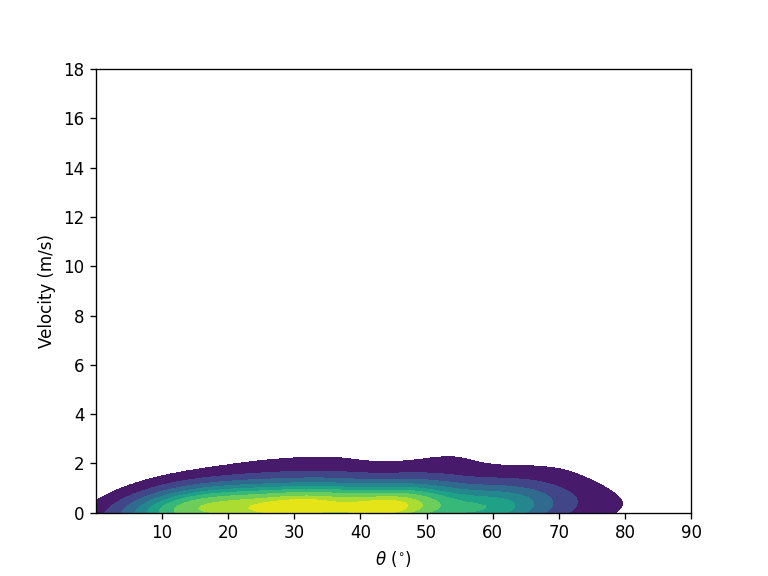}
    \includegraphics[width=0.49\columnwidth]{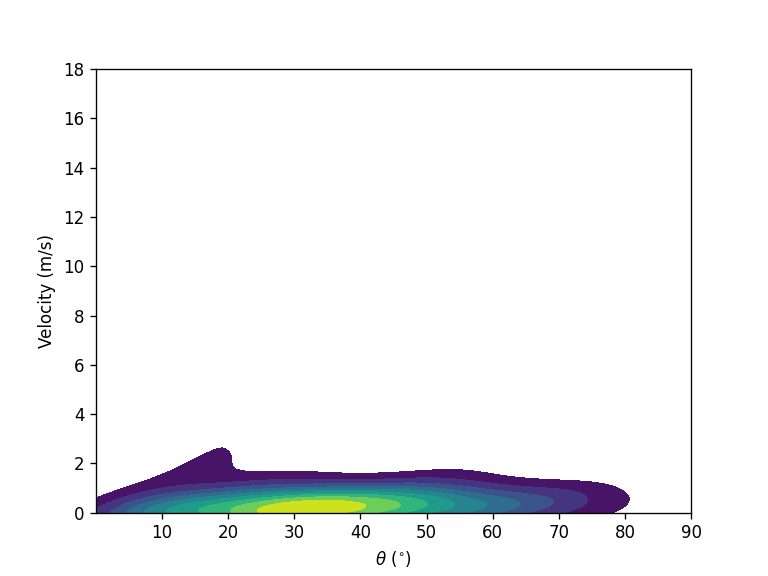}
    \includegraphics[width=0.49\columnwidth]{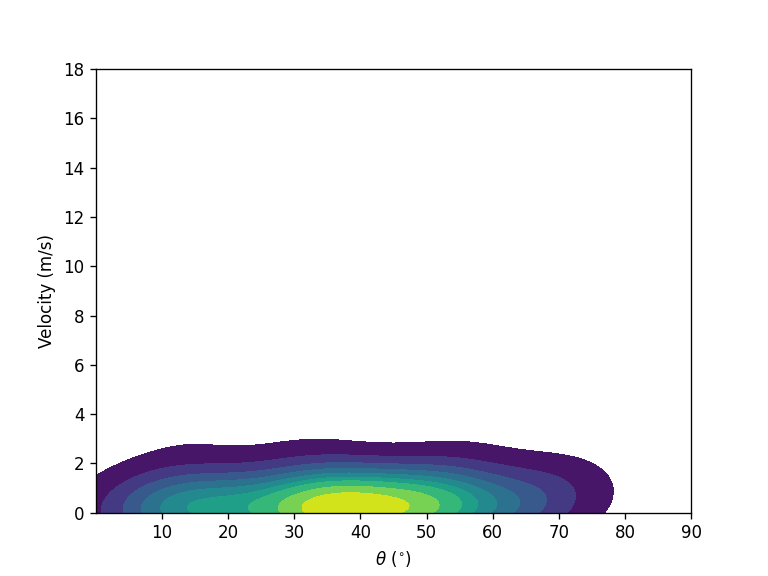}
    \caption{Relative velocities as a function of the impact angle ($\theta$) in collisions
    between two macroscopic bodies. The upper panel represents simulations using a density of
    0.5 g/cm$^3$, while the bottom uses 0.9 g/cm$^3$. The left plots include Aegaeon
    in the simulations, whereas the right ones exclude the satellite. These plots illustrate
    Aegaeon's influence on the velocity and impact angle distribution in collisions among
    macroscopic bodies.  The color
    map ranges from less likely (blue) to more likely (yellow).}
    \label{fig:collsions_macroscopic}
\end{figure}

Figure \ref{fig:collsions_macroscopic} shows a heatmap of relative velocity
versus impact angle for collisions between two 20 m radius bodies. Upper panels
represent simulations with 0.5 g/cm$^3$ density, lower panels with 0.9 g/cm$^3$.
Left plots include Aegaeon, right plots exclude it. Results indicate typical
relative velocities of approximately 0.5 m/s, with velocities up to 10 m/s
depending on body location. Impact angles are predominantly below 60$^\circ$,
typically around 30$^\circ$. Aegaeon's presence significantly influences the
distribution of velocities and impact angles.

The velocity distribution appears more uniform in simulations excluding Aegaeon,
indicating its significant effect on body velocities within the arc, especially
during close encounters. Typical collision velocities between two macroscopic
bodies (0.5 m/s) differ notably from those involving Aegaeon (5 m/s),
demonstrating Aegaeon's role in accelerating nearby bodies. The acceleration
caused by Aegaeon to a body approximately 500 m away is on the order of
$10^{-4}$ m/s$^2$, compared to Saturn's acceleration of about $1$ m/s$^2$ in the
arc.

The next section presents the Smoothed Particle Hydrodynamics (SPH) simulations of the most probable impact scenarios
derived from our N-body simulations. These simulations focus on collisions
between macroscopic bodies and impacts involving Aegaeon, aiming to quantify
dust production within the G-ring arc.

%%%%%%%%%%%%%%%%%%%%%%%%%%%%%%%%%%%%%%%%%%%%%%%%%%
%SPH
\section{Smoothed particle hydrodynamics simulations - SPH}
\label{section:sph}

In realistic collision scenarios, the resultant fragment distribution is non-uniform,
exhibiting significant variation in both size and velocity. To accurately understand
the physical processes at play, it is essential to account for these variations. While
the literature offers semi-analytical models for estimating collision outcomes and
dust production, such as those proposed in \cite{leinhardt2012a}, these models primarily
focus on collisions between bodies in the gravitational regime. Our study, however,
deals with collisions involving smaller bodies, which fall outside the gravitational
regime. Consequently, these semi-analytical models are unsuitable for our purposes,
making it necessary to simulate collisions to comprehend the dynamics beyond the usual
regime.

In this study, we utilize parameters derived from N-body simulations as inputs for SPH
simulations. This approach enables us to compute the mass, velocity, and dust
production resulting from the collisions. Our methodology involves employing a hybrid
hydrocode for simulating the initial disruption of small bodies at the collision point.
Subsequent SPH simulations will be conducted to investigate shock propagation, material
modification, and gravitational reaccumulation processes.

\subsection{Physical model}

For modeling the collisional process and accurately predicting its outcome and dust
production, we utilized a GPU-accelerated SPH code \textsc{miluphcuda} \cite{Schaefer2016,
schafer2020versatile}. This code implements a hydrodynamic approach combined with
fracture modeling. The code solves the standard conservation equations of mass, momentum, and energy,
which are pivotal in depicting the behavior of elastic solids. This functionality enables
precise modeling of the deformation and fragmentation of colliding bodies. We
incorporated a plasticity model, utilizing an extended Von Mises-Drucker-Prager yield
criterion \cite{fish_1997}, which modifies stresses beyond the elastic limit. This aspect
is crucial for reflecting the material strength of the bodies and ensuring realistic
simulation outcomes.

For brittle failure and fragmentation, we applied the approach by \cite{Benz1999} along
with the \cite{Grady1980} model. In this model, the material flaws, distributed according
to a Weibull distribution, act as points of weakness that can become active and grow under
tensile stress. The parameters for this distribution, specifically for ice, were derived by
fitting data to rate-dependent fracture stress measurements in \cite{Lange1984}.

Under higher strain rates, materials experience gradual separation due to crack
development and flaw activation, with cracks growing at a constant velocity. The effect of
fractures is quantified through a scalar state variable, the damage, which represents the
local stress reduction resulting from crack propagation. This variable serves as an
indicator of the material's reduced strength under tensile loading. Upon reaching a fully
damaged state, the material transforms into a cohesionless fluid, thus losing its capacity
to withstand damage.

We employed the Tillotson equation of state (EOS) \cite{Tillotson1962} to simulate bodies
under extreme conditions. This EOS is computationally efficient and adaptable, suitable
for simulations involving a large number of SPH particles or complex geometries. It effectively
represents various physical states, including high temperatures, pressures, and densities,
making it versatile for astrophysical simulations.

%%%%%%%%%%%%%%%%%%%%%%%%%%%%%%%%%%%%%%%
\subsection{SPH simulations}

In the SPH simulations, we focused on generating dust 
from collisions between macroscopic bodies under various conditions. 
We examined how changes in body radii influence fragility and 
the collisional disruption of meter-size objects. 
Additionally, we explored velocities from high-impact to lower thresholds, 
distinguishing between cohesive collisions and mere adhesion.

A typical collision setup involves two bodies of equal size colliding at a fixed impact angle
of $\theta = 30^\circ$. The target is impacted by a projectile moving at velocity $V_i$.
The impact angle $\theta$ is determined at initial contact, formed by the line connecting
their centers and the projectile's trajectory. The radii of the target and projectile, $R$
and $r$, respectively, influence collision dynamics. The impact parameter,
$b = \sin \theta = B/(R + r)$, is calculated based on the impact geometry, where $B$ is
the distance from the center of the target to the first contact point. The length $l$
represents the projectile's overlap with the target.

To initialize the SPH simulations, the target and projectile were positioned at a mutual
distance of $r_{ini} = R + r + 2.5h$, with $h$ denoting the smoothing length, defining the
region of influence for neighboring SPH particles.

Fragment identification, dynamics analysis, composition study, and dust production from a
collision are intricate processes. We divided these analyses into two primary steps:
First, a 'friends-of-friends' algorithm identifies spatially connected SPH particle clusters
as "fragments." Then, gravitationally bound "aggregates" of these fragments are determined.
The largest aggregate is initially identified using the most massive fragment and all
gravitationally bound fragments. An iterative approach refines this aggregate. The process
repeats to identify the second-largest aggregate, starting with the most massive fragment not
yet included in the largest one \cite{Burger2017}.

The simulations used 150,000 SPH particles for each body, ensuring sufficient resolution for
various collision scenarios. The runtime of the simulations depended on the body's size and
the impact velocity, which determined the overall simulated time.

We determine dust production based on the SPH particle resolution in the simulation. 
Dust predominantly consists of fragments representing less than 0.01\% of the total mass, 
composed of single fully damaged SPH particles. 
By aggregating the mass of fragments below this threshold, 
we quantify the dust generated post-collision. 
These estimates should be considered lower bounds, 
as they do not account for potential enhancement from surface regolith 
layers that could be present on the bodies. 
The presence of fine-grained surface material could increase dust 
generation efficiency, particularly in low-velocity impacts, 
though quantifying this effect would require additional assumptions 
about poorly constrained regolith properties.

Table \ref{tab:dust_generation} summarizes dust production from collisions between two
same-size bodies, considering their radii and impact velocities, with a uniform impact angle of
30 degrees for all scenarios.  At 2 m/s, no dust is generated regardless
of body size. As the velocity increases, dust generation rises notably, especially for
larger bodies. For instance, at 10 m/s, a small amount of dust is produced for bodies with
radii of 5 m and 10 m, but not for smaller or larger ones, suggesting a threshold velocity
for dust generation relative to body size. 
Final snapshots of these simulations, showing the outcomes,
are illustrated in Figure \ref{fig: Snapshots tabela} in the appendix.

\begin{table}[!ht]
	\centering
	\caption{Dust generation from different collision setups using a fixed impact angle of 30$^\circ$.}
	\label{tab:dust_generation}
	\begin{tabular}{llllll} % five columns, alignment for each
        Radius & & 1 m & 5 m & 10 m & 20 m\\ 
        \hline 
        & Velocity & \multicolumn{4}{c}{Dust produced} \\
	\hline
        & 2 m/s  & 0.00\%  & 0.00\%  & 0.00\%  & 0.00\% \\
        & 10 m/s & 0.00\%  & 0.14\%  & 0.34\%  & 2.13\% \\
        & 20 m/s & 0.01\%  & 0.62\%  & 0.60\%  & 2.01\% \\ 
        \hline
	\end{tabular}
\end{table}

\begin{figure}[!h]
\centering
\begin{subfigure}[t]{0.43\linewidth}
\includegraphics[width=\linewidth]{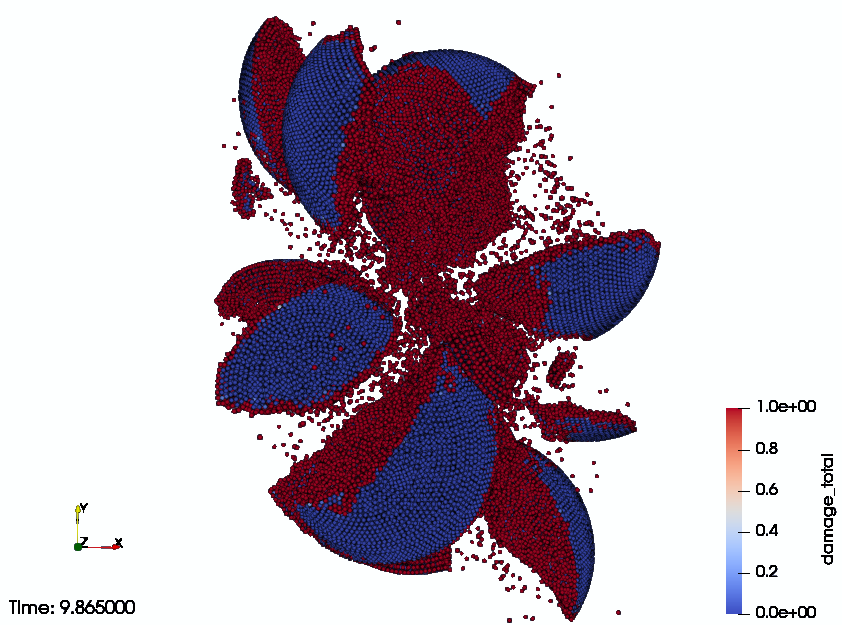}
\caption{Snapshot showing the damaged SPH particles.}
\end{subfigure}
\quad
\begin{subfigure}[t]{0.48\linewidth}
\includegraphics[width=\linewidth]{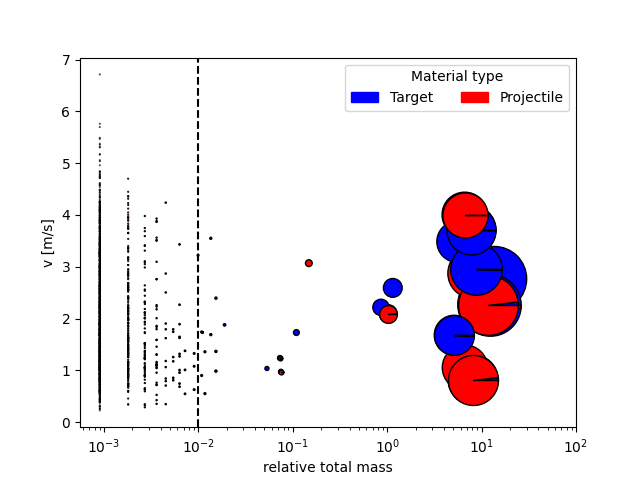}
\caption{Relative mass of fragments and their velocity.}
\end{subfigure}
\caption{Fragments and dust from an impact at 10 m/s.}
\label{fig: damage and v mass 09}
\end{figure}

Figure \ref{fig: damage and v mass 09} visually distinguishes between damaged SPH particles
(left panel) and the mass fragments' distribution along with their velocities. This
visualization clearly identifies the fragments resulting from the simulation.

Collisions among macroscopic bodies generally occur at low velocities, leading to minimal
dust production regardless of the bodies' sizes. To investigate alternate dust generation
mechanisms, we focused on collisions with Aegaeon, constituting about 40\% of total collisions
in the arc verified in the N-body simulations.

\begin{figure}[h!]
\centering
\begin{subfigure}[t]{0.43\linewidth}
\includegraphics[width=\linewidth]{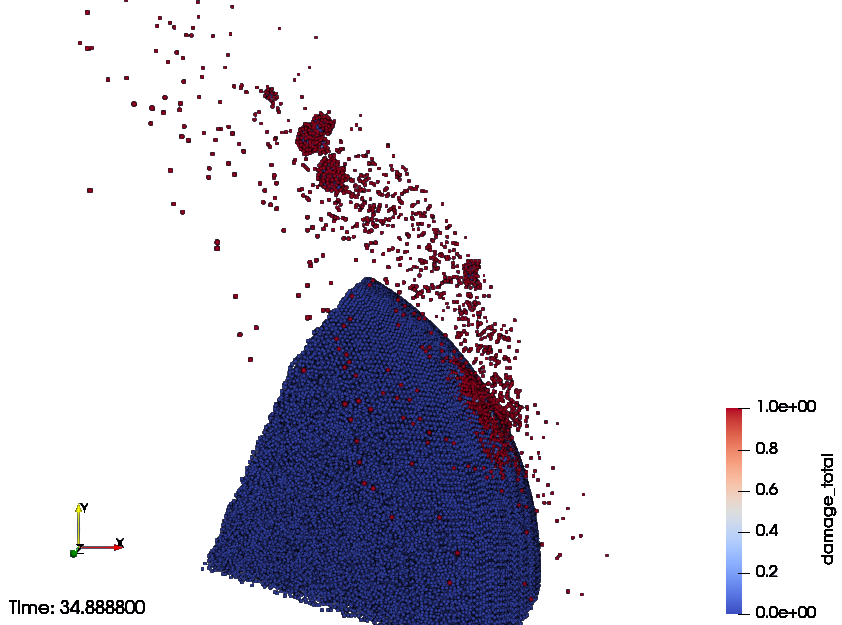}
\caption{Snapshot showing damaged SPH particles from the projectile impacting Aegaeon.}
\end{subfigure}
\quad
\begin{subfigure}[t]{0.48\linewidth}
\includegraphics[width=\linewidth]{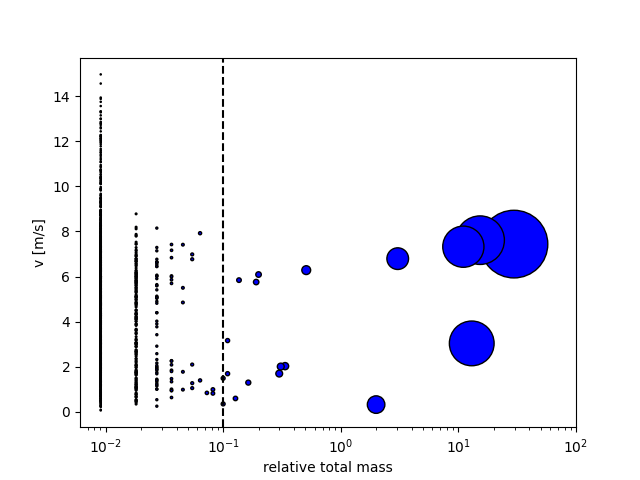}
\caption{Velocity and mass distribution of fragments.}
\end{subfigure}
\caption{Fragments and dust from a 10 m/s impact.}
\label{fig: Aeg r20 v10 theta 45}
\end{figure}

We adopted a standard collision scenario involving a relative velocity between 5 m/s and 10 m/s
and an impact angle of 45$^\circ$. This setup aligns with typical collisions observed in our
simulations. Figure \ref{fig: Aeg r20 v10 theta 45} illustrates the mass fragment distribution
and corresponding velocities from an SPH simulation of a collision between a 240 m-sized target
(Aegaeon) and a 20 m-sized projectile at 10 m/s, at an angle of 45$^\circ$. The projectile
was simulated with 10,000 SPH particles, while Aegaeon was modeled using 1,000,000 SPH particles.
Due to the resolution and the size of the bodies, only a portion of the satellite was included in
the simulation. The results show significant disruption of the projectile, with 76.47\% of its mass
forming smaller fragments, and the remaining 23.53\% becoming dust.

Table \ref{tab: Aeg e project} details dust production resulting from various collision
scenarios between a projectile and Aegaeon, with different projectile velocities
and impact angles.The dust production varies significantly with
changes in the projectile's velocity and impact radius. In typical collisions with Aegaeon,
moderate velocities and impact angles tend to result in an average dust production of around 10\%.

\begin{table}[!ht]
\centering
\caption{Dust production from a collision between a projectile with Aegaeon.}
\begin{tabular}{llclcl}
\hline
            & Velocity & \multicolumn{4}{c}{Dust produced}                                                                                   \\ \hline
Projectile: &          & \multicolumn{1}{l}{r =10 m} & \multicolumn{1}{l|}{$\theta=30^\circ$} & \multicolumn{1}{l}{r = 20 m} & $\theta=45^\circ$ \\ \hline
            & 5 m/s    & \multicolumn{2}{c|}{9.38\%}                                          & \multicolumn{2}{c}{6.63\%}                   \\
            & 10 m/s   & \multicolumn{2}{c|}{17.37\%}                                         & \multicolumn{2}{c}{23.53\%}                  \\ \hline
\end{tabular}
\label{tab: Aeg e project}
\end{table}

A more comprehensive discussion of the results from all SPH simulations will be
addressed in a forthcoming paper, with the distribution of
fragments resulting from the collisions and estimate the proportion of these
fragments that remain embedded in the arc post-collision.

%%%%%%%%%%%%%%%%%%%%%%%%%%%%%%%%%%%%%%%%%%%%%%%%%%%%%
%Dust Production
\section{Dust production through impacts with Aegaeon}
\label{section-dustproduction}

In simulations of various impact setups with Aegaeon, it
was observed that approximately 10-70\% of the impactor's mass converts into dust
and smaller fragments.

\begin{figure}[!ht]
\centering
\includegraphics[width=0.7\columnwidth]{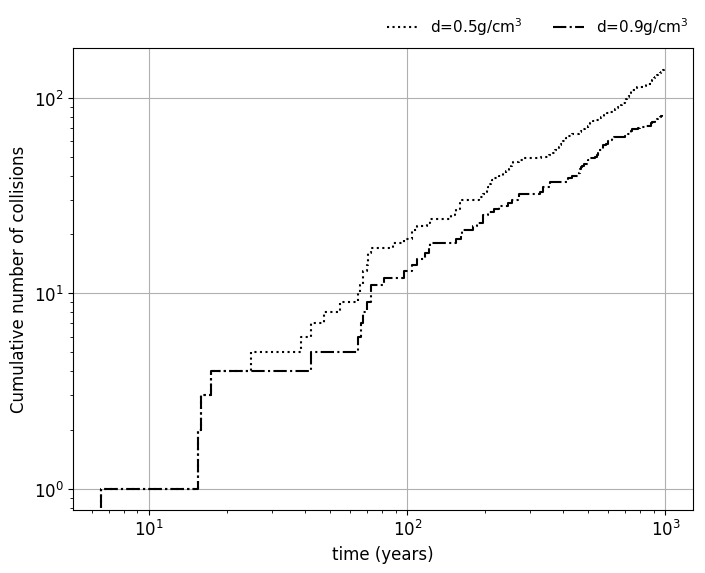}
\caption{Cumulative number of collisions with the satellite Aegaeon for bodies
with densities of 0.5 g/cm$^3$ and 0.9 g/cm$^3$.}
\label{fig:cumulative:aeg}
\end{figure}

Figure \ref{fig:cumulative:aeg} shows the total number of impacts of macroscopic bodies on
Aegaeon over time, based on N-body simulations. The full-dashed line represents simulations
using bodies with a density of 0.5 g/cm$^3$, and the dashed line for those with 0.9 g/cm$^3$.
In the simulation with porous bodies, there were 142 impacts, while in the non-porous
simulation, there were 83 impacts over 1000 years. From this data, we can roughly estimate
an impact frequency with Aegaeon at about one collision every 8 years for porous bodies and
one every 13 years for non-porous bodies, based on a simple linear fit. However, this ratio
is only applicable over a short period, as the bodies gradually disintegrate over time.
Additionally, the cumulative effect of producing fragments in each collision, which may
subsequently collide with Aegaeon, generating more dust, is not considered in this calculation.

To estimate and assess the evolution of dust in the arc and examine scenarios that could
keep the dust population in a steady state, we applied an exponential decay model based on
the dust particles' lifetime \cite{madeira_2018}. Our analysis spans various particle sizes,
from 1$\mu$m to 10$\mu$m. Additionally, every time a collision with Aegaeon was detected in
our simulations, we added the corresponding amount of dust generated by that collision into
the system. This approach allowed us to integrate the continuous production of dust from
these impacts. By combining the natural decay of dust particles with the new dust generated
from collisions with Aegaeon, we could compute the dust population within the arc.

\begin{figure}[!ht]
\centering
\includegraphics[width=0.84\columnwidth]{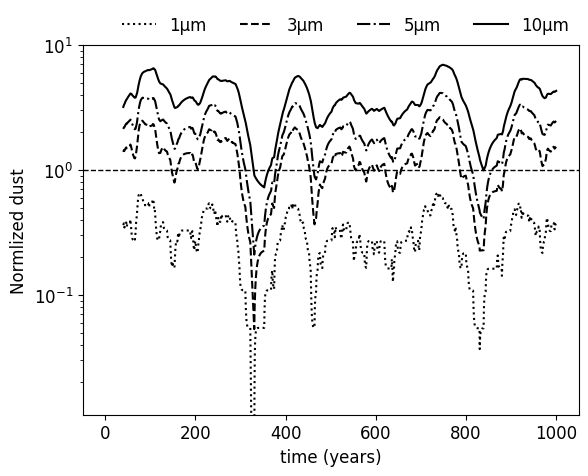}
\includegraphics[width=0.84\columnwidth]{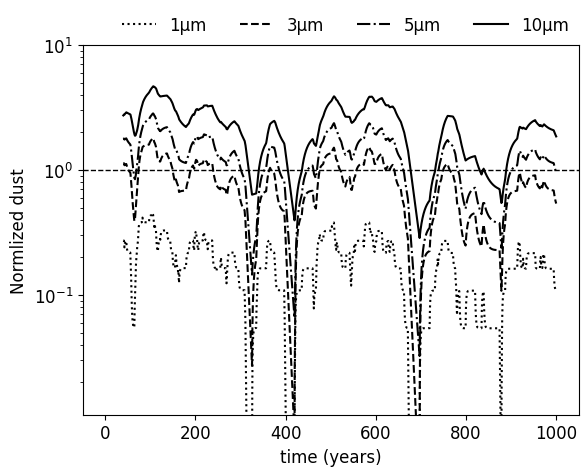}
\caption{Normalized dust quantity (by the current dust amount in the arc) over time.
The plots represent dust produced by collisions between 20 m radius bodies and
Aegaeon, interpolated with decay due to solar pressure radiation for dust particles
of $1\mu$m, $3\mu$m, $5\mu$m, and $10\mu$m. The upper panel is for collisions
with 0.5 g/cm$^3$ density bodies, and the bottom for 0.9 g/cm$^3$. The dashed line
represents the current dust amount in the arc, approximately $10^6$ kg.}
\label{fig:dust_evolution}
\end{figure}

Figure \ref{fig:dust_evolution} shows the moving average of dust quantity (normalized by the
current dust amount in the arc, approximately $10^6$ kg) over time for each scenario at
$v=5v_{esc}$. The evolution patterns for dust particles at $v=1v_{esc}$ and $v=10v_{esc}$
show similar general behaviors, as the focus is on the moving average of dust quantity.

Our findings suggest it is possible to maintain dust population in the arc for the simulated
period of 1000 years, both for the porous or non-porous scenarios. However, keeping 1$\mu$m
dust particles is challenging due to their short lifespan. We then propose that the arc's dust could
originate from collisions between macroscopic bodies and Aegaeon, leading to their
fragmentation and dust generation. This hypothesis becomes more compelling when considering
an underestimation of collision frequency with Aegaeon, as we only accounted for 20 m radius
bodies. A more realistic scenario with bodies ranging from 1 to 50 m in radius could
significantly increase collision rates and dust production. Such a scenario supports the theory
that the G-ring arc may have originated from a super-catastrophic collision, fragmenting a
larger body into Aegaeon (the largest fragment) and numerous smaller fragments, which continue
to generate dust in the arc.

The impacts studied can effectively replenish the dust particles within the arc, and the arc
lifetime depends on the quantity of bodies immersed in it. Under the examined scenario, a
balanced system could be sustained for at least a thousand years as these bodies gradually
fragment. Nonetheless, this system is likely to be transitory.

Estimating the distribution of these bodies in terms of radius is challenging, but it is clear
that a higher number of bodies would prolong the arc's existence. Moreover, our impact rate
estimation with Aegaeon is conservative, as we have not included in the analysis the likelihood
of subsequent collisions involving fragments generated from initial impacts.

Additionally, we might consider the possibility of a 'firefly' behavior characterized by
fluctuating brightness. In a situation with few bodies in the arc, the frequency of impacts
with Aegaeon would reduce significantly, leading to brief periods of intensified brightness
following each collision. This hypothesis highlights the dynamic nature of the arc's brightness
based on the population and activity of macroscopic bodies within it.

%%%%%%%%%%%%%%%%%%%%%%%%%%%%%%%%%%%%%%%%%%%%%%%%%%%%%

\section{Mass evolution of Aegaeon}
\label{section-massevolution}
We also examined the mass evolution of Aegaeon through its collisions with the bodies within
the arc. Across all simulations, Aegaeon did not experience any catastrophic collisions, nor
did it lose a significant amount of mass. The satellite sustained minimal damage and remained
intact in all types of collisions modeled in the N-body simulations. These findings imply that
the macroscopic bodies, being at least three orders of magnitude less massive than Aegaeon,
can coexist with the satellite without substantially affecting its mass. According to
\cite{hedman_2010}, these bodies are believed to be no larger than 50 meters in diameter,
insufficient to cause significant destruction to Aegaeon, given the relative velocities
observed in the arc.

For completeness, we also simulated the arc with 2000 larger bodies, ranging from 50 to 100
meters, under the assumption of perfect merging to potentially create a second body comparable
in mass to Aegaeon. Two sets of simulations were conducted: one including Aegaeon and the other
without it. These simulations revealed that when Aegaeon was present, it grew in size but prevented
other bodies from similarly increasing in mass. Conversely, in the simulation without Aegaeon, a
single body rapidly gained mass while inhibiting growth in others. These outcomes suggest that
the formation of other massive bodies through collisions is unlikely, as Aegaeon or another
dominant body tends to block the mass accumulation in other bodies within the arc.

%%%%%%%%%%%%%%%%%%%%%%%%%%%%%%%%%%%%%%%%%%%%%%%%%%%%%%
%Discussion

%%%%%%%%%%%%%%%%%%%%%%%%%%%%%%%%%%%%%%%%%%%%%%%%%%%%%

\section{Final Remarks}
\label{section-finalremarks}
The G-ring arc is predominantly composed of micrometric dust particles, which are greatly
influenced by external forces, particularly solar radiation pressure. This pressure
results in the removal of these dust particles from the arc in under 40 years. Aegaeon,
the satellite, cannot replenish these dust particles quickly enough through collisions
with interplanetary particles, as it requires at least 3000 years to generate the
current dust mass. Consequently, an alternative mechanism is necessary to keep
the dust population in the arc.

One possible mechanism involves the potential existence of macroscopic bodies in the arc. Collisions
between these bodies, as well as with Aegaeon, could be a source of dust. This study
investigated this theory through N-body and SPH simulations, focusing on the dust
generated by such collisions. The findings indicate that collisions among macroscopic
bodies alone are insufficient to replenish the arc's dust. However, impacts involving
Aegaeon can convert 10-60\% of the impactor mass into dust, depending on the collision
parameters. A standard collision was assumed in this study, with a relative velocity
of 5 m/s and an impact angle of 45$^\circ$, and it was posited that each impact with
Aegaeon could convert 10\% of the impactor's mass into dust, aiding in arc
replenishment.

Our results demonstrate that the arc can survive for thousands of years due to
impacts with Aegaeon. These impacts do not lead to significant mass loss for Aegaeon,
which remains intact post-collision. The study modeled the macroscopic body population
as exclusively comprising 20 m bodies. From the SPH collision simulations, it is apparent
that even a single impact is sufficient to repopulate the arc's dust material.
However, as these bodies gradually break apart, the material for generating new dust
diminishes, and the impact frequency decreases. Thus, the arc's lifespan depends on
the distribution and total number of macroscopic bodies present.

The study did not account for the possibility that each collision 
could produce additional fragments that may collide again with Aegaeon, 
potentially extending the arc's lifespan through a collisional cascade. 
While our simulations suggest impact frequencies of one collision every 8-13 years 
depending on body density, these rates could increase over time as 
fragments from previous collisions become new impactors. 
This cascade effect, combined with resonant confinement of collision 
products, could sustain the arc beyond the thousand-year timescale 
examined here. Precisely estimating the arc's ultimate lifespan is 
challenging due to uncertainties in fragment size distribution and resonant capture efficiency of collision products.

The G-ring arc may be temporary and exhibits fluctuating dust levels,
dependent on the number and size distribution of macroscopic bodies. Precisely
estimating the arc's lifespan is challenging due to these variables. Nonetheless,
the presence of these bodies, if confirmed, could be crucial for maintaining the
arc through collisions with Aegaeon. A small number of bodies in the arc could lead
to intermittent brightness spikes following impacts, whereas a larger population
might result in a more stable and consistently bright system. A large
population would also allow the arc to be long-lived instead of some thousands of years. 

In a
more general sense, this mechanism can also be applied to other arc systems where a swarm of
macroscopic bodies may be embedded. Future work will address this point and 
involve additional SPH simulations to gain deeper insights into dust production
and the trajectories and velocities of the fragments generated by these impacts.

%\ack{We thank to ....}

\dataccess{The data is available through contact with the authors.}

\funding{This work was supported by the following entities: \\
CAPES - Coordenação de Aperfeiçoamento de Pessoa de Nível Superior - Process Number: 1684418 \\
CAPES -  Coordenação de Aperfeiçoamento de Pessoal de Nível Superior - Brasil (CAPES) - Finance Code 001 \\
DFG German Research Foundation project 446102036 \\
FAPESP (Processo 2016/24561-0)
}

\conflict{The authors declare no conflict.}

%%%%%%%%%% Insert bibliography here %%%%%%%%%%%%%%
 \bibliographystyle{RS}
  \bibliography{references}

\newpage
\begin{appendices}
    \section{Simulation Snapshots}

    \begin{figure*}
        \vspace*{-20pt}
          \centering
          \begin{subfigure}{0.7\linewidth}
            \caption{r = 1m}
            \includegraphics[width=\linewidth]{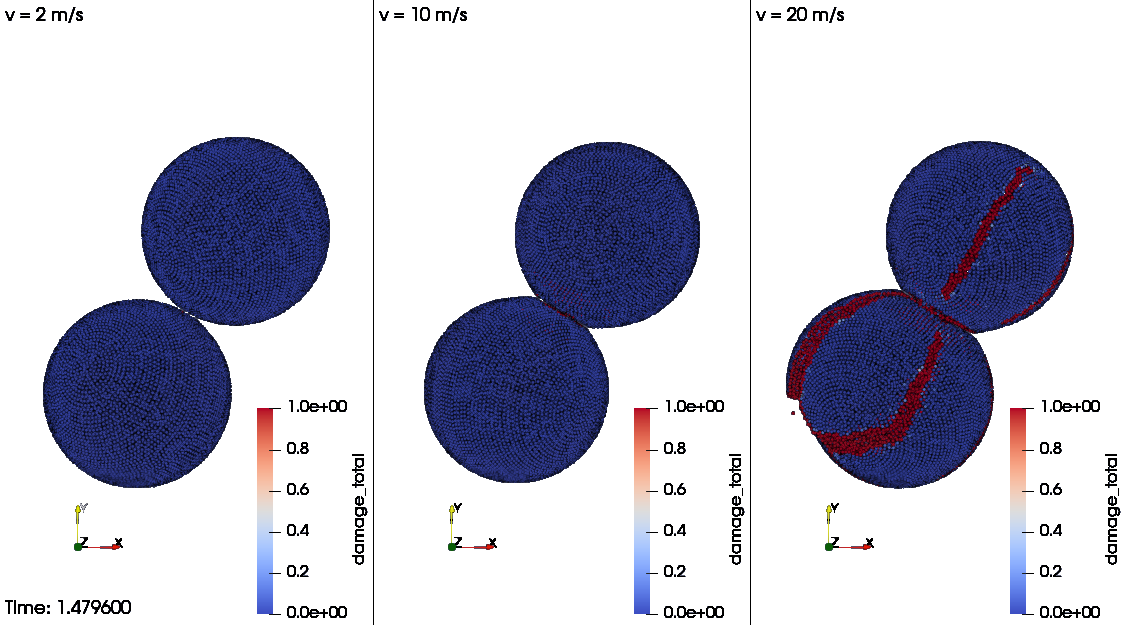}
          \end{subfigure}
          \begin{subfigure}{0.7\linewidth}
            \caption{r = 5m}
            \includegraphics[width=\linewidth]{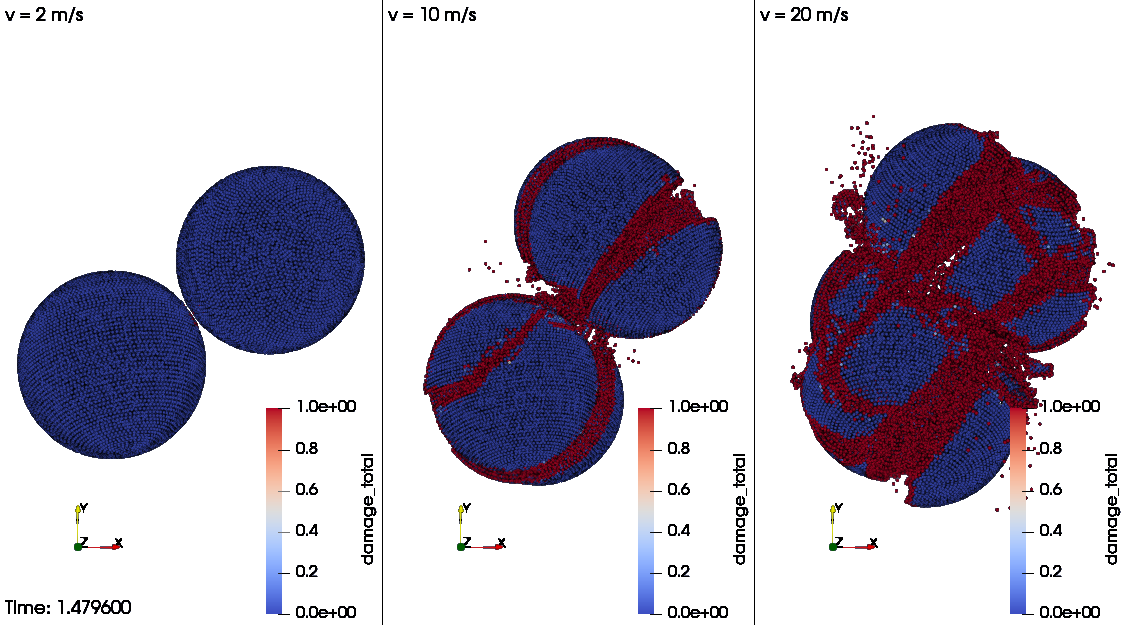}
          \end{subfigure}
          \begin{subfigure}{0.7\linewidth}
            \caption{r = 10m}
            \includegraphics[width=\linewidth]{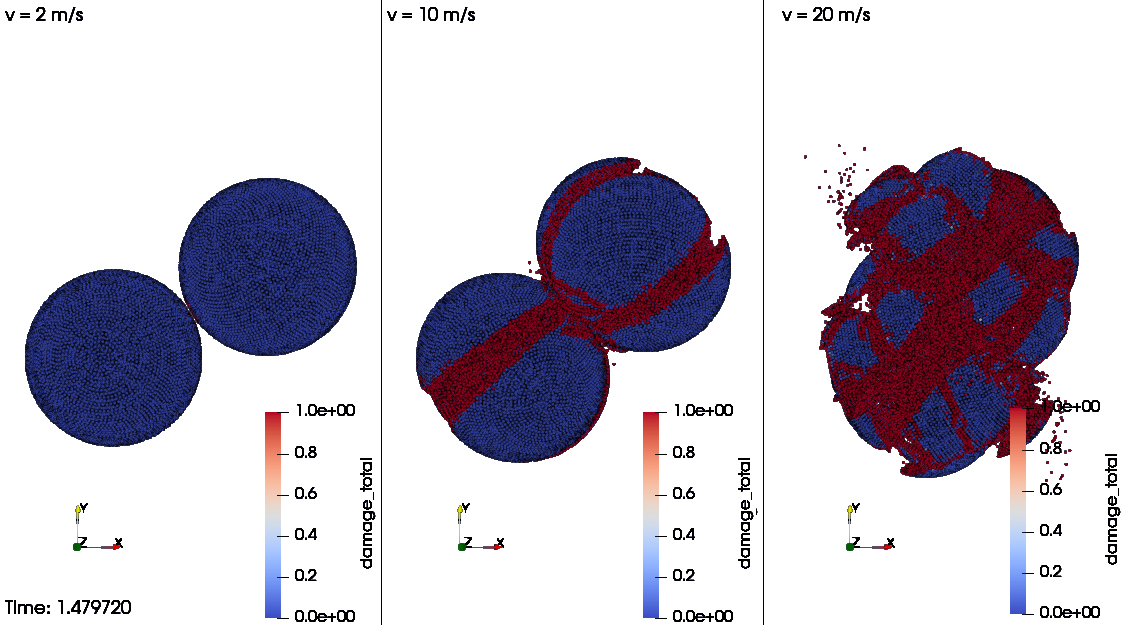}
          \end{subfigure}
          \begin{subfigure}{0.7\linewidth}
            \caption{r = 20m}
            \includegraphics[width=\linewidth]{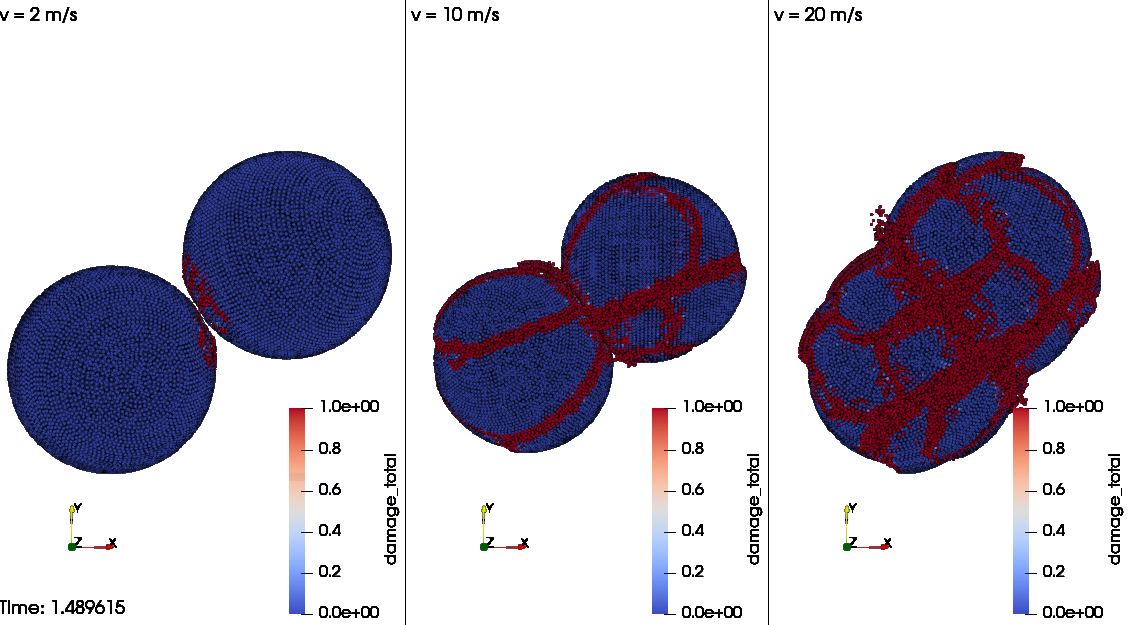}
          \end{subfigure}
            \caption{Snapshots from different collision setups using a fixed impact angle ($\theta=30^\circ$). }
          \label{fig: Snapshots tabela}
    \end{figure*}

    \begin{figure}[h!]
          \centering
          \begin{subfigure}[t]{0.4\linewidth}
            \includegraphics[width=\linewidth]{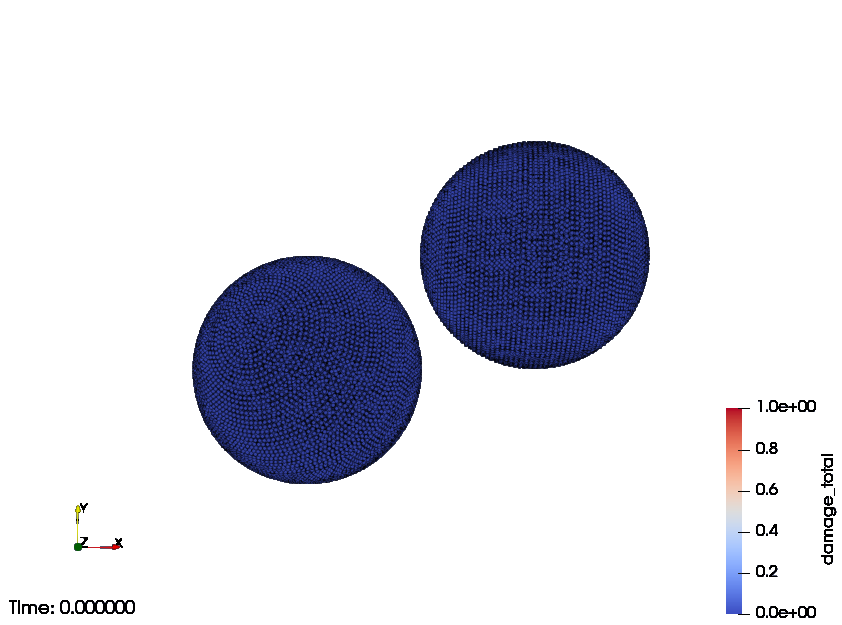}
          \end{subfigure}
          \begin{subfigure}[t]{0.4\linewidth}
            \includegraphics[width=\linewidth]{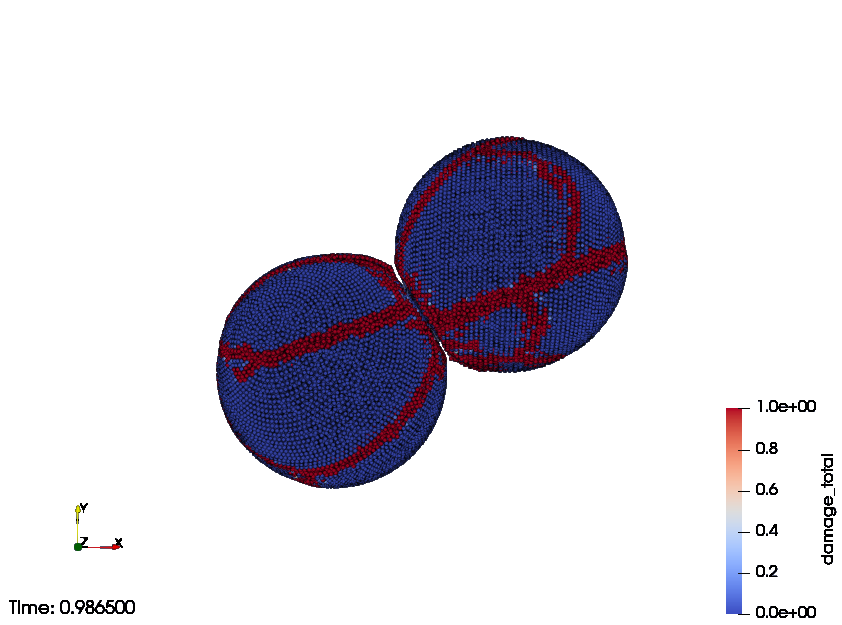}
          \end{subfigure}
          \begin{subfigure}[c]{0.4\linewidth}
            \includegraphics[width=\linewidth]{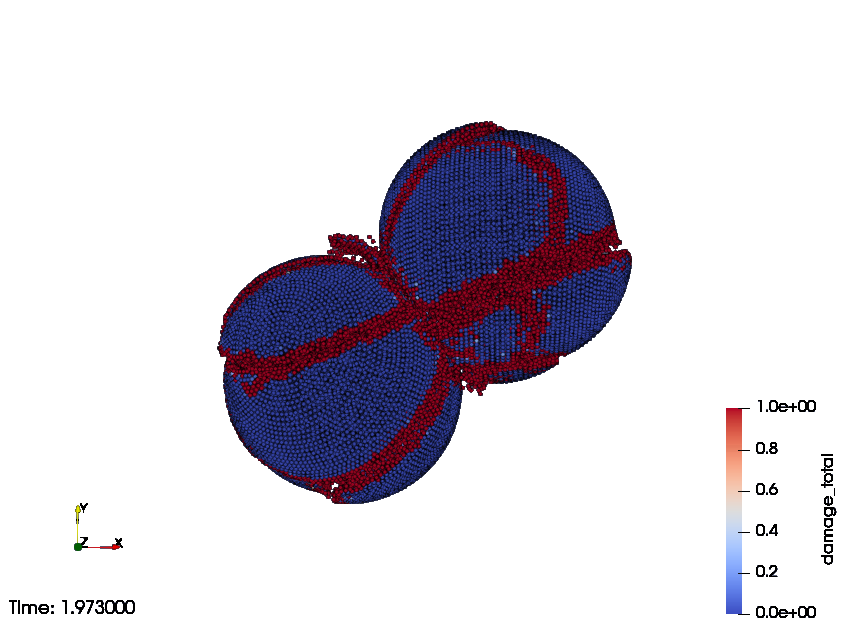}
          \end{subfigure}
          \begin{subfigure}[c]{0.4\linewidth}
            \includegraphics[width=\linewidth]{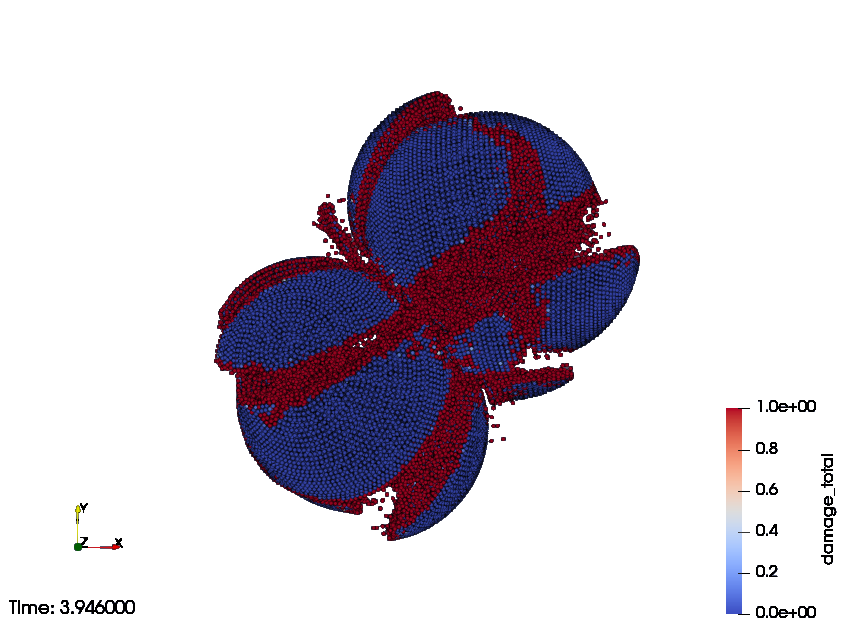}
          \end{subfigure}
          \begin{subfigure}[b]{0.4\linewidth}
            \includegraphics[width=\linewidth]{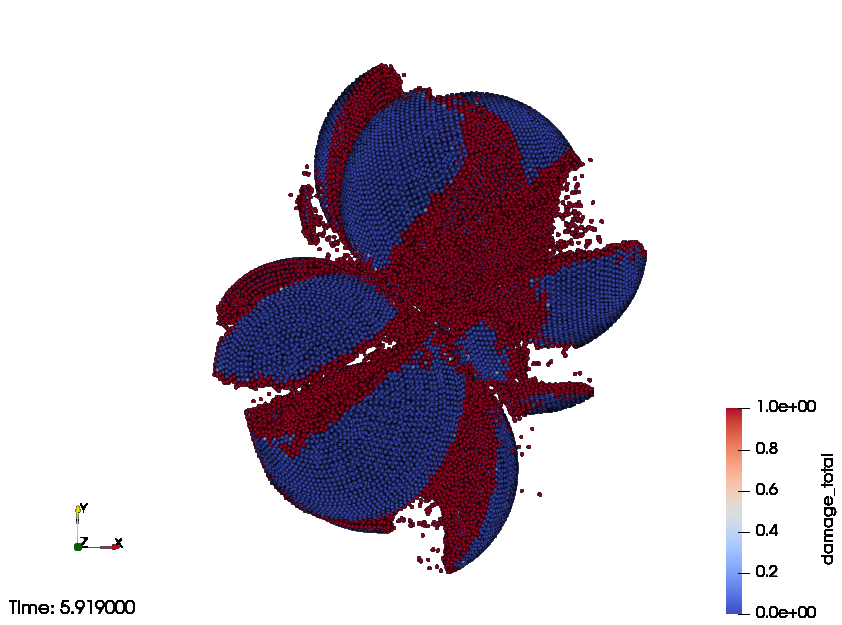}
          \end{subfigure}
          \begin{subfigure}[b]{0.4\linewidth}
            \includegraphics[width=\linewidth]{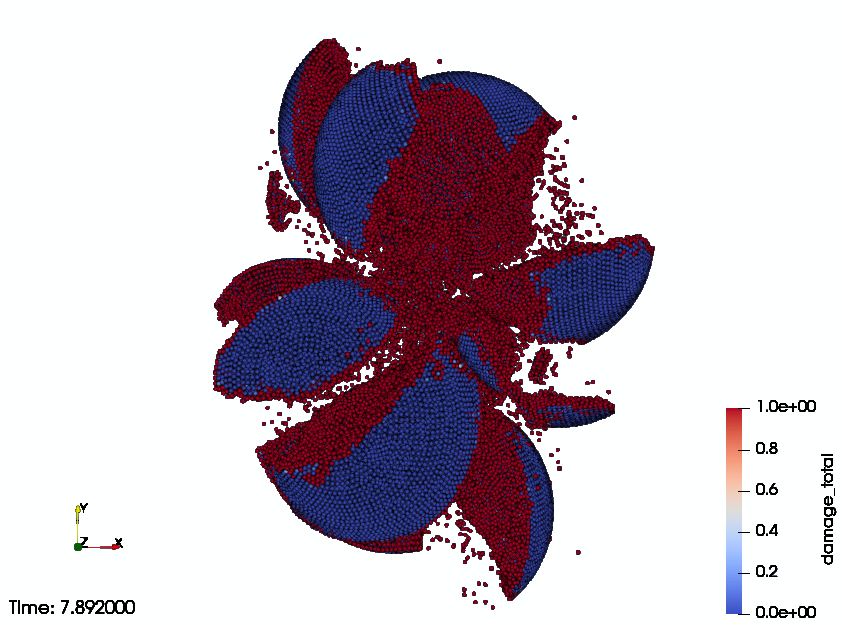}
          \end{subfigure}
        \begin{subfigure}[b]{0.4\linewidth}
            \includegraphics[width=\linewidth]{figures/r20_v10/r20_v10_sp.0500-branco.png}
          \end{subfigure}
          \begin{subfigure}[b]{0.4\linewidth}
            \includegraphics[width=\linewidth]{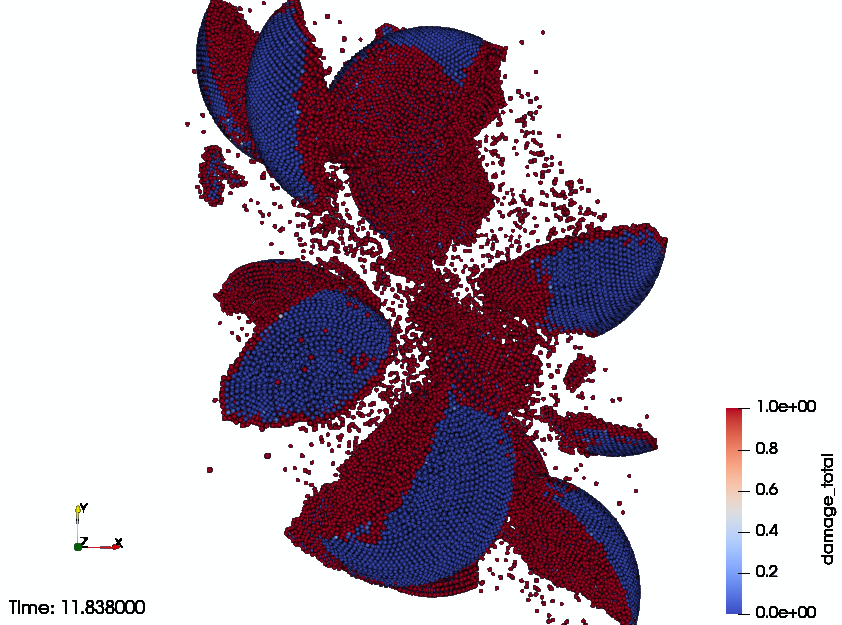}
          \end{subfigure}
          \caption{Snapshots showing the damage from two equal 20 m-sized bodies at v = 10 m/s and $\theta=30^\circ$ angle of impact. }
          \label{fig: Snapshot r 20 v 10}
    \end{figure}

    \begin{figure}[h!]
          \centering
          \begin{subfigure}[t]{0.4\linewidth}
            \includegraphics[width=\linewidth]{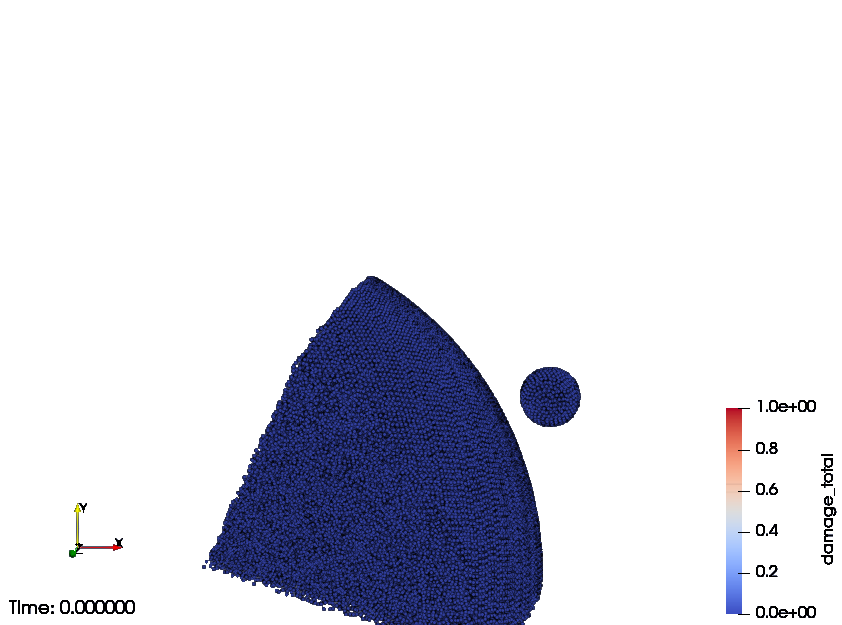}
          \end{subfigure}
          \begin{subfigure}[t]{0.4\linewidth}
            \includegraphics[width=\linewidth]{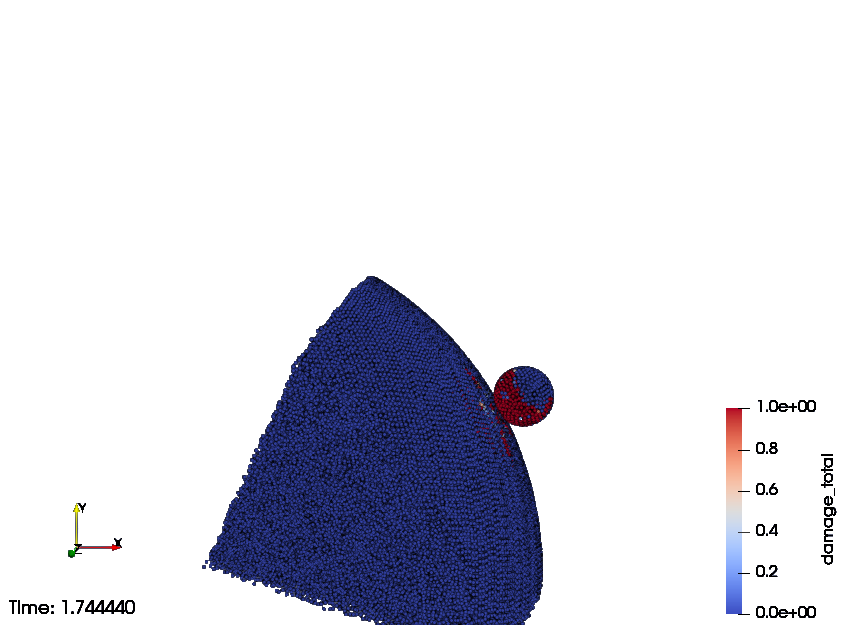}
          \end{subfigure}
          \begin{subfigure}[c]{0.4\linewidth}
            \includegraphics[width=\linewidth]{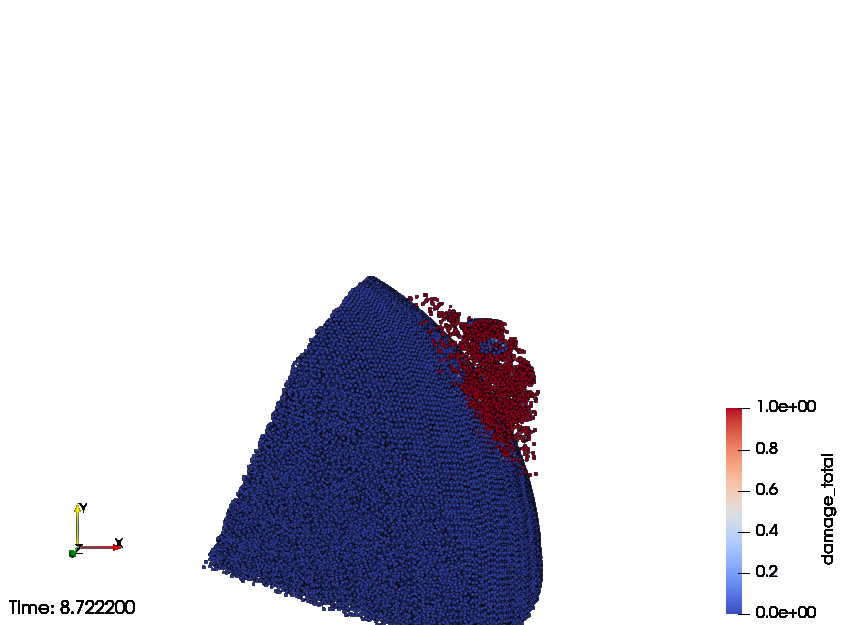}
          \end{subfigure}
          \begin{subfigure}[c]{0.4\linewidth}
            \includegraphics[width=\linewidth]{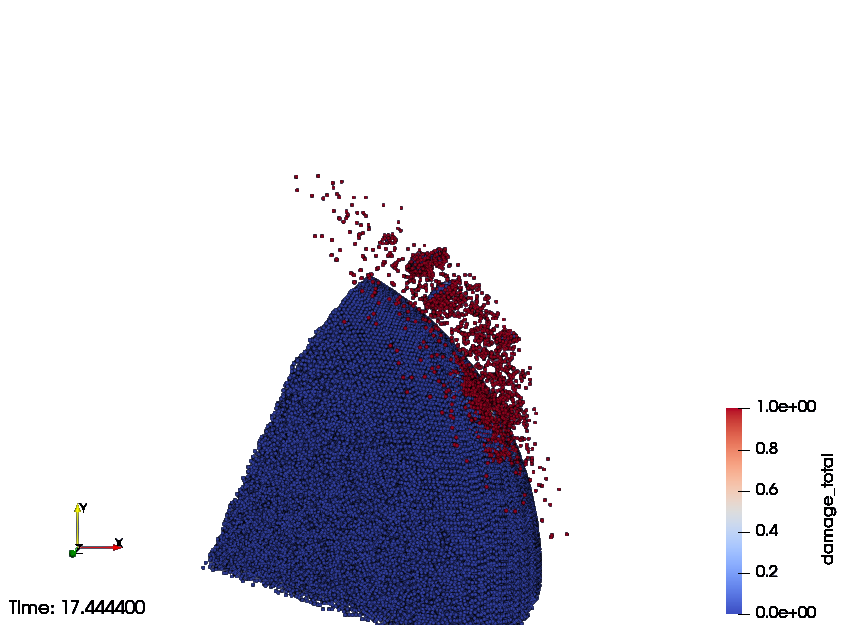}
          \end{subfigure}
          \begin{subfigure}[b]{0.4\linewidth}
            \includegraphics[width=\linewidth]{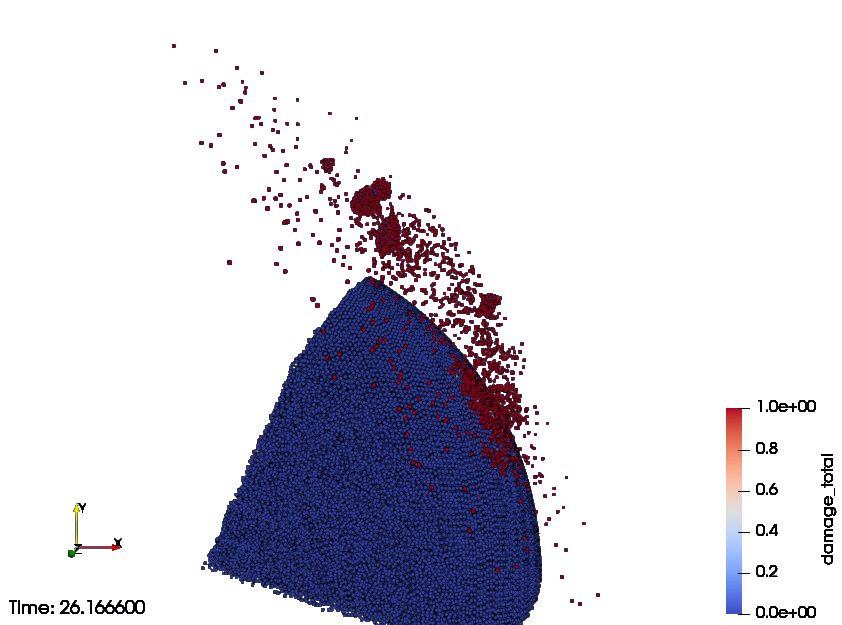}
          \end{subfigure}
          \begin{subfigure}[b]{0.4\linewidth}
            \includegraphics[width=\linewidth]{figures/Aeg_r20_v10_theta45/Aeg_r10_v10_t45.0200-branco.png}
          \end{subfigure}
        \begin{subfigure}[b]{0.4\linewidth}
            \includegraphics[width=\linewidth]{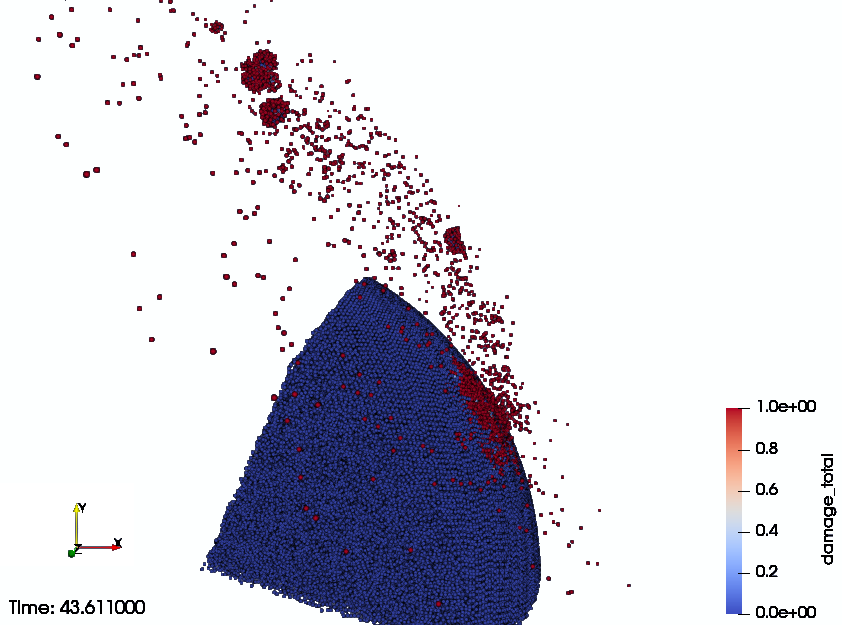}
          \end{subfigure}
          \begin{subfigure}[b]{0.4\linewidth}
            \includegraphics[width=\linewidth]{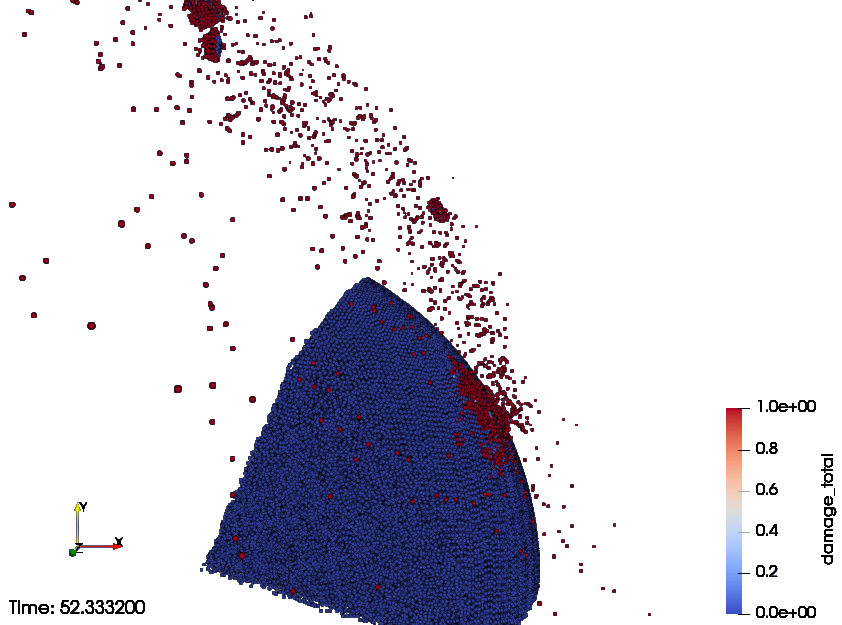}
          \end{subfigure}
          \caption{Snapshots of Aegeon impacted by a 20 m-sized projectile at 10 m/s and an angle of $45^\circ$. }
          \label{fig: Snapshot r 20 v 10}
    \end{figure}

\end{appendices}

\end{document}